\documentstyle[12pt,epsf]{article}
\let\section=\subsection     \let\subsection=\subsubsection                

\begin{document}
\begin{flushright}
  LBNL-45192
\end{flushright}
\ \\
\begin{center}
   {\large \bf DILEPTONS FROM TRANSPORT AND}\\[2mm]
   {\large \bf HYDRODYNAMICAL MODELS}\\[5mm]
   P.~HUOVINEN and V.~KOCH \\[5mm]
   {\small \it  Lawrence Berkeley National Laboratory (LBNL) \\
                1 Cyclotron Road, Berkeley CA 94720, USA \\[8mm] }
\end{center}

\begin{abstract}\noindent
  Transport and hydrodynamical models used to describe the expansion stage
  of a heavy-ion collision at the CERN SPS give different dilepton spectrum
  even if they are tuned to reproduce the observed hadron spectra. To
  understand the origin of this difference we compare the dilepton emission
  from transport and hydrodynamical models using similar initial states in
  both models. We find that the requirement of pion number conservation in a
  hydrodynamical model does not change the dilepton emission. Also the mass
  distribution from the transport model indicates faster cooling and
  longer lifetime of the fireball.
\end{abstract}

\section{Introduction}

Electromagnetic signals from relativistic heavy-ion collision directly probe
the properties of the dense matter created during the collision since their
interactions with the surrounding matter are negligible. However, the observed
lepton pairs and photons do not originate only at one temperature and density,
but the distribution is a complicated integral over the entire space-time
history of the system. Therefore, to draw any conclusions of the observed
yield, one has to understand the evolution of the system and how it affects
dilepton emission.

The evolution of the system is a complicated many-body problem which can not be
solved from basic principles but has to be described using phenomenological
models instead. Various models based on hydrodynamics and transport theory have
been successfully used to describe the hadron data measured in A+A collisions
at the CERN SPS energies. However, when they are used to describe dilepton
emission in the same collisions, the dilepton yields around invariant mass 500
MeV differ roughly by a factor two~\cite{Huovinen99,K}. At this mass region the
CERES collaboration at CERN has measured a significant excess of dileptons over
the estimated background~\cite{Lenkeit}. It has been suggested that this
enhancement might be an in medium effect or possibly a precursor of chiral
symmetry restoration~\cite{Rapp99us}, but before drawing any such conclusions
one has to understand why different expansion dynamics can lead to equally
large enhancements.

To investigate the effect of expansion dynamics to dilepton production
we have compared the dilepton yields from three different models --
transport, hydrodynamical model with zero pion chemical potential and
hydrodynamical model with conserved pion number.

\section{The models}

To simplify the study of expansion dynamics we have kept the particle content
of the system as simple as possible. The only particles included are pions
and rho mesons and the only production channel for electron pairs is
$\pi\pi$ annihilation. No in-medium modifications of particle properties
have been taken into account, but all cross sections, widths etc.\ are those
of free particles.

The transport model we use is the relativistic BUU transport model described in
ref.~\cite{Koch96} and the hydrodynamical model the 2+1 dimensional non-boost
invariant model described in ref.~\cite{Sollfrank97}. One of the important
differences between these models is that pion number is conserved in the
transport model but not in the hydrodynamic model. The pion number conservation
leads to non-zero pion chemical potential which is one of the possible causes
of the difference in the dilepton yields~\cite{Rapp99}. To study the effect of
non-zero pion chemical potential in the framework of a hydrodynamical expansion
we made a new version of the hydrodynamic model where the conserved baryon
number is replaced by a conserved pion
number\footnote{We define the conserved pion number as
                ${\cal N}_{\pi} = n_{\pi} + 2 n_{\rho}$, where $n_{\pi}$ and
                $n_{\rho}$ are the actual number densities of pions and
                rho-mesons respectively.}.

As mentioned the only dilepton production channel we consider is
$\pi\pi$ annihilation. The cross section for this process used in
the transport description is given in ref.~\cite{Koch96} whereas
the thermal production rate used in the hydrodynamical description
is the one calculated by Gale and Kapusta~\cite{Gale87}.

We have checked the consistency of our calculations by imposing periodic
boundary conditions to our models, initializing the systems in thermal
and chemical equilibrium and checking that the equilibrium is maintained.
In this case the dilepton emission from all three models is identical
and corresponds to the thermal rate at this temperature.

In the simulations of the actual heavy-ion collisions, the initial
state of the evolution is chosen to reproduce the observed hadron
spectra~\cite{Huovinen99,Koch96}. However, in the present calculations
we use the same initial state for all models: a spherical fireball with
a radius of $r=8$ fm in thermal and chemical equilibrium with no initial
flow. The density profile is assumed to be Woods-Saxon with the maximum
energy density of $\epsilon = 0.5$ GeV/fm$^3$ which corresponds to a
maximum initial temperature of $T=218$ MeV, pion number density
$n_{\pi}=0.38$~fm$^{-3}$ and rho number density $n_{\rho} = 0.20$~fm$^{-3}$.
Initially the system contains 560 pions and 260 rhos. The edge of the system
is defined by the radius where temperature drops below the decoupling
temperature of the hydrodynamic model. This temperature is set to be
$T_{dec} = 120$ MeV in both versions of the hydrodynamic model whereas
there is no need for a decoupling temperature in the transport model. In
the pion number conserving hydro decoupling at $T_{dec} = 120$ MeV leads
to an average pion chemical potential on the decoupling surface of
$\langle\mu_{\pi}\rangle = 75$ MeV.

\section{Results}

\begin{figure}
\begin{center}
  \begin{minipage}[t]{6.7cm}
        \epsfysize 6.8cm \epsfbox{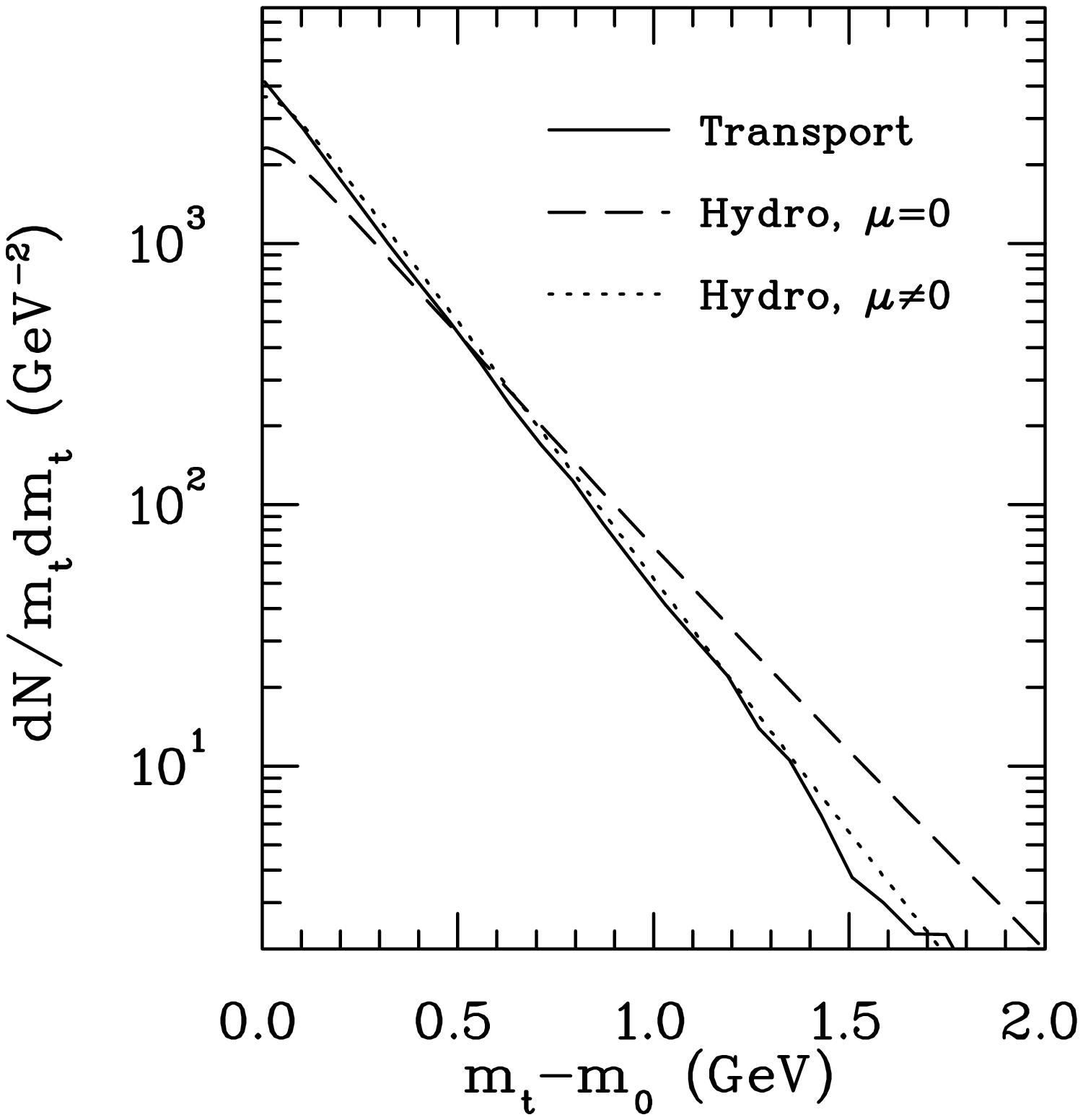}
        \hfill
    \caption{The $p_t$ spectra of pions from transport and hydrodynamical
             models.}
    \label{pions}
  \end{minipage}
   \hfill
  \begin{minipage}[t]{6.7cm}
        \epsfysize 6.8cm \epsfbox{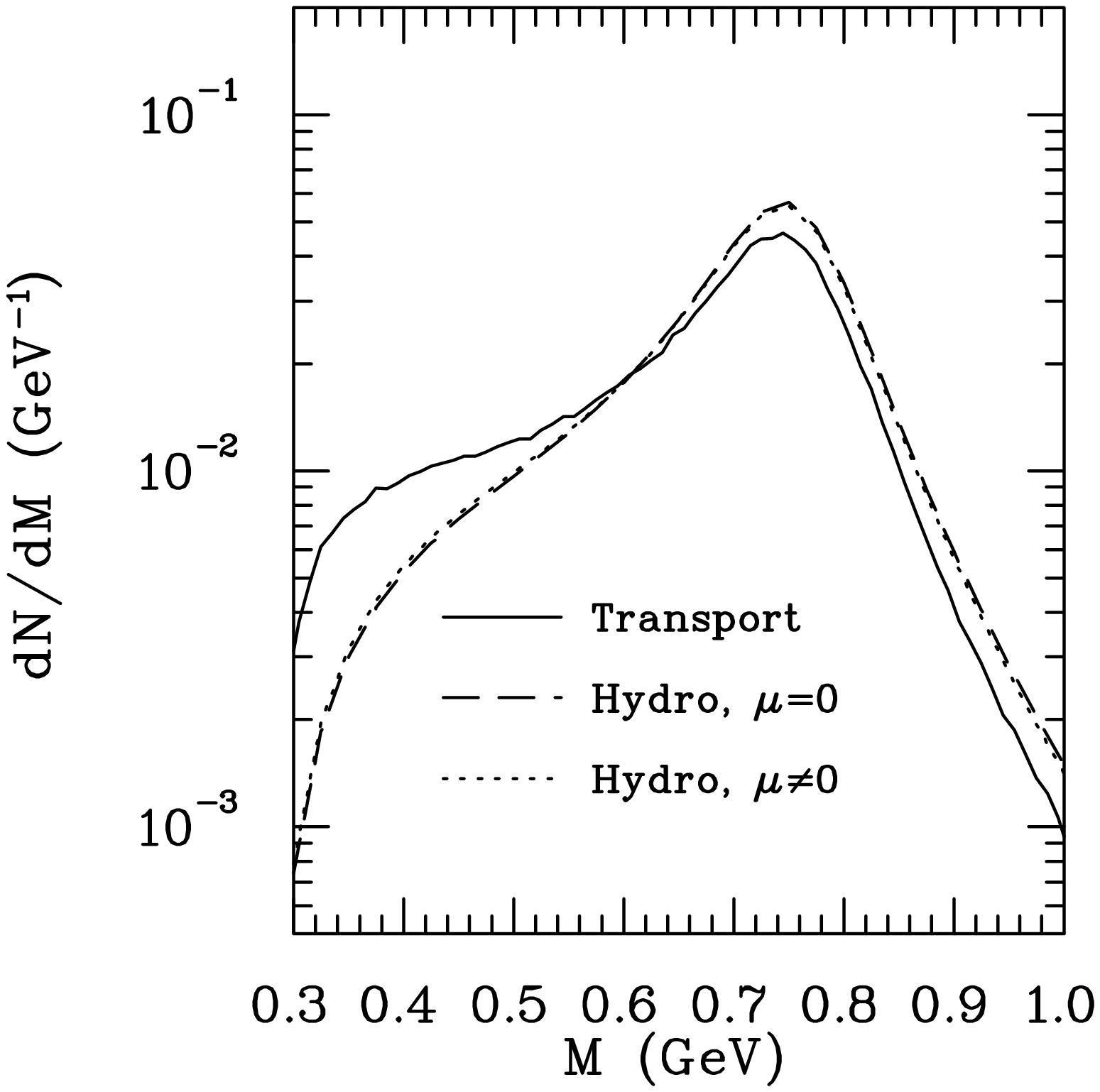}
        \hfill
    \caption{The mass distribution of lepton pairs from $\pi\pi$ annihilation
             in transport and hydrodynamical models.}
    \label{leptons}
  \end{minipage}
\end{center}
\end{figure}

Since the simulations of the actual heavy-ion collisions are tuned to
reproduce the observed hadron spectra, we calculate the pion spectra
as well. The resulting $p_t$ spectra of pions is shown in
fig.~\ref{pions}. In the hydrodynamic model with zero chemical potential
the pion number is not conserved and the number of final pions is smaller
than in the other two models. Another difference is that the effective
equation of state of the transport model is softer than in the hydrodynamic
model. This is manifested in the slope of the $p_t$ spectrum which is steeper
for transport calculation than for hydrodynamic calculation with zero
chemical potential.

The pion number conserving hydro gives almost similar $p_t$ slope compared to
transport. The steeper slope than in zero chemical potential hydro is easily
understood. When the system dilutes and pion number is conserved, a larger
fraction of energy is stored in the mass of pions than in the case of zero
chemical potential. This leads to faster decrease of temperature and the
decoupling temperature is reached at an earlier stage of evolution when the
flow is less developed.

Fig.~\ref{leptons} depicts the distribution of lepton pairs originating from
$\pi\pi$ annihilations during the system evolution. The most striking feature
is that the difference between the two hydrodynamical models is tiny. The
effect of increasing chemical potential and thus larger pion density is
counterbalanced by the shorter lifetime and faster cooling of the system
leading to practically indistinguishable dilepton yields. However, the pion
spectra from these two models are different. If the models are required to
produce similar pion spectra, the initial state of the model with zero chemical
potential should be larger and have lower initial temperature than the pion
number conserving model. This difference in the initial state would also lead
to different dilepton production.

Since the transport model and the hydrodynamic model lead to similar pion
spectra their dilepton yields can be compared without reservations. The
difference between these models is similar to that seen in the attempts to
reproduce the CERES data. This supports our hypothesis that details of
expansion dynamics do have a significant effect on the dilepton production.
The shapes of the distributions look like the system in transport description
cools faster but lives longer than in hydro. Whether this is the case remains
to be investigated in more detail.

We have demonstrated that the effect of the expansion dynamics on dilepton
production is visible and that the non-zero pion chemical potential is not the
main cause of this effect. At the present stage of the work there are still
many open questions like the temperature evolution in the transport description
and when and where the dileptons are emitted. It also has to be checked how the
distributions change if all the models are required to produce similar pion
spectra.

\section*{Acknowledgements}

This work was supported by the Director, 
Office of Science, Office of High Energy and Nuclear Physics, 
Division of Nuclear Physics, and by the Office of Basic Energy
Sciences, Division of Nuclear Sciences, of the U.S. Department of Energy 
under Contract No. DE-AC03-76SF00098.

\end{document}